\documentclass{aip-cp}

\usepackage[numbers]{natbib}
\usepackage{rotating}
\usepackage{graphicx}
\usepackage{url}
\usepackage[utf8]{inputenc}

\begin{document}

\newcommand{\gadf}{\texttt{gamma-astro-data-formats} }
\newcommand{\gadfgithub}{\url{https://github.com/open-gamma-ray-astro/gamma-astro-data-formats} }
\newcommand{\gadfrtd}{\url{https://gamma-astro-data-formats.readthedocs.io/} }

\newcommand{\ogragithub}{\url{https://github.com/open-gamma-ray-astro} }
\newcommand{\ogralist}{\url{https://lists.nasa.gov/mailman/listinfo/open-gamma-ray-astro} }
\newcommand{\ogrameudon}{\url{https://github.com/open-gamma-ray-astro/2016-04_IACT_DL3_Meeting/} }

\newcommand{\fermipy}{\url{http://fermipy.readthedocs.io/} }
\newcommand{\fermist}{\url{http://fermi.gsfc.nasa.gov/ssc/data/analysis/software/} }
\newcommand{\pointlikedata}{\url{https://github.com/tburnett/Fermi-LAT/blob/master/pointlike_document/Data\%20Format.ipynb} }

\newcommand{\ogip}{\url{https://heasarc.gsfc.nasa.gov/docs/heasarc/ofwg/ofwg_intro.html} }

\title{Open high-level data formats and software for gamma-ray astronomy}
\corresp[cor1]{Corresponding author: Christoph.Deil@mpi-hd.mpg.de}
\corresp[cor2]{Corresponding author: catherine.boisson@obspm.fr}

\author[mpik]{Christoph Deil\corref{cor1}}
\author[obsparis]{Catherine Boisson\corref{cor2}}
\author[cea]{Karl Kosack}
\author[jeremy]{Jeremy Perkins}
\author[mpik]{Johannes King}
\author[mpik]{Peter Eger}
\author[humboldt]{Michael Mayer}
\author[matthew]{Matthew Wood}
\author[victor]{Victor Zabalza}
\author[irap]{Jürgen Knödlseder}
\author[ifae-bist]{Tarek Hassan}
\author[erlangen]{Lars Mohrmann}
\author[erlangen]{Alexander Ziegler}
\author[apc]{Bruno Khelifi}
\author[erlangen]{Daniela Dorner}
\author[desy]{Gernot Maier}
\author[desy]{Giovanna Pedaletti}
\author[ifae-bist]{Jaime Rosado}
\author[ifae-bist]{José Luis Contreras}
\author[obsparis]{Julien Lefaucheur}
\author[dortmund]{Kai Brügge}
\author[obsparis]{Mathieu Servillat}
\author[apc]{Régis Terrier}
\author[roland]{Roland Walter}
\author[saverio]{Saverio Lombardi}

\affil[mpik]{MPIK, Heidelberg, Germany}
\affil[dortmund]{TU, Dortmund, Germany}
\affil[cea]{CEA/IRFU/SAp, CEA Saclay, Bat 709, Orme des Merisiers, 91191 Gif-sur-Yvette, France}
\affil[jeremy]{NASA/GSFC, USA}
\affil[obsparis]{LUTH, Observatoire de Paris, Meudon, France}
\affil[apc]{APC, Universit\'e Paris Diderot, CNRS/IN2P3, Paris, France}
\affil[erlangen]{FAU, Erlangen, Germany}
\affil[humboldt]{Humboldt University, Berlin, Germany}
\affil[desy]{DESY, Zeuthen, Germany}
\affil[roland]{Observatoire de Genève, 51 chemin des Maillettes, 1290 Sauverny, Switzerland}
\affil[gae-ucm]{Universidad Complutense de Madrid}
\affil[ifae-bist]{Institut de Fisica d’Altes Energies (IFAE), The Barcelona Institute of Science and Technology, Campus UAB, 08193 Bellaterra (Barcelona) Spain}
\affil[irap]{IRAP, Toulouse, France}
\affil[jeremy]{NASA/GSFC}
\affil[matthew]{SLAC National Accelerator Laboratory}
\affil[saverio]{INAF, Osservatorio Astronomico di Roma, via Frascati 33, 00040 Monte Porzio Catone (Roma), Italy}
\affil[victor]{University of Leicester, UK}

\maketitle

\begin{abstract}
In gamma-ray astronomy, a variety of data formats and proprietary software have been traditionally used, often developed for one specific mission or experiment. Especially for ground-based imaging atmospheric Cherenkov telescopes (IACTs), data and software are mostly private to the collaborations operating the telescopes. However, there is a general movement in science towards the use of open data and software. In addition, the next-generation IACT instrument, the Cherenkov Telescope Array (CTA), will be operated as an open observatory.

We have created a Github organisation at \ogragithub where we are developing high-level data format specifications. A public mailing list was set up at \ogralist and a first face-to-face meeting on the IACT high-level data model and formats took place in April 2016 in Meudon (France). This open multi-mission effort will help to accelerate the development of open data formats and open-source software for gamma-ray astronomy, leading to synergies in the development of analysis codes and eventually better scientific results (reproducible, multi-mission).

This write-up presents this effort for the first time, explaining the motivation and context, the available resources and process we use, as well as the status and planned next steps for the data format specifications. We hope that it will stimulate feedback and future contributions from the gamma-ray astronomy community.

\end{abstract}

\newpage

\section{Introduction}

The Flexible Image Transport System (FITS) format was created around 1980 \cite{Wells:1981} by optical astronomers. In the 1990s, the HEASARC FITS Working Group, also known as the OGIP (Office of Guest Investigator Programs) FITS Working Group, produced documents and recommendations concerning the storage of X-ray (and partly gamma-ray space telescope) data in FITS.\footnote{\ogip} Several of these recommendations have subsequently been incorporated into the FITS standard, the latest version is FITS 3.0 from 2010 \cite{Pence:2010}.

\begin{figure}[tb]
\centerline{\includegraphics[width=1\textwidth]{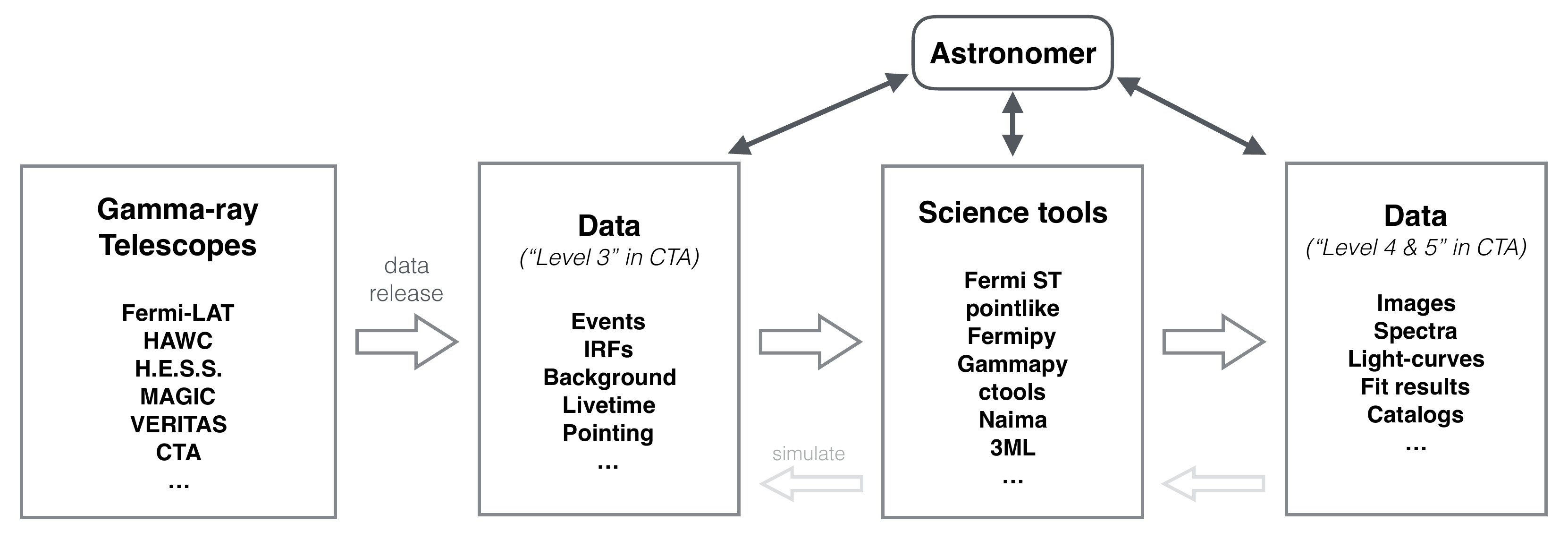}}
\caption{
The purpose of the \texttt{gamma-astro-data-formats} effort is to encourage collaboration between high-level gamma-ray data producers, science tool developers, and data analysts. The goal is to develop common data formats to avoid duplication of efforts and confusion by astronomers working with multi-mission gamma-ray data or multiple analysis tools.
}
\label{fig:purpose}
\end{figure}

Today, very-high energy (VHE, energy~$>$~50~GeV) gamma-ray astronomy is finding itself in a similar situation like X-ray astronomy in the 1990s (illustrated in Figure~\ref{fig:purpose}). The existing ground-based imaging atmospheric Cherenkov telescopes (IACTs) like e.g. H.E.S.S., MAGIC and VERITAS, have been operating independently for the past decade, using proprietary data formats and codes. Data from each IACT is stored in ROOT files containing serialised C++ objects and can only be read with the private software. The Cherenkov Telescope Array (CTA), the next generation IACT instrument, will be operated as an an open observatory, meaning that data and analysis software will be public to all astronomers. Current IACTs have started to ``export'' their data and instrument response functions (IRFs) to FITS, partly as a prototyping effort for CTA, but also to take advantage of the open-source science tool codes for gamma-ray astronomy (Gammapy \cite{2015arXiv150907408D}, ctools \cite{2016AnA...593A...1K}, pointlike \citep{2010PhDT.......147K}, Fermi ScienceTools\footnote{\fermist}, Fermipy\footnote{\fermipy}, 3ML \citep{2015arXiv150708343V}, Naima \citep{2015arXiv150903319Z}, \ldots) and to have an archival and common data data format that allows joint analysis with other astronommical multi-wavelength datasets.
For science data products, the term ``data level 3'' (DL3) is used for event lists, IRFs and auxiliary data for analysis and provenance, ``data level 4'' (DL4) for higher-level science data products like images, spectra and lightcurves, and ``data level 5'' (DL5) for source catalogs (see Figure~\ref{fig:purpose}).

This situation (many gamma-ray data producers and science tools) has prompted us to start in early 2016 the \gadf effort -- an attempt to create an open forum and process to create gamma-ray data models and formats. In some cases we are using or extending the existing formats (mainly FITS and OGIP recommendations), in some cases we are creating new formats that more directly reflect our use cases. The goal is to improve collaboration between people working on this topic and to produce data format specifications to help data producers, tool developers, and astronomers working with high-level gamma-ray data.

\newpage

\section{Resources, Process, Work Product}

The goal of the \gadf effort is to enable efficient collaboration on gamma-ray data formats and codes. To this end, we have set up the following resources that are open to anyone interested in the topic:

\begin{itemize}
\item{} A mailing list (currently 75 members, including people from all major gamma-ray collaborations) with this official description: ``This group is organized for the discussion of software and data formats for the gamma-ray astronomy community. If you are interested in open and common data and software formats for space- and ground-based instruments you are encouraged to join.'': \\ \ogralist
\item{}A Github organisation for online collaboration on data format specifications via issues and pull requests:\\ \gadfgithub
\item{}Our main work product, the data format specifications, are available online at:\\ \gadfrtd
\item{}We hold monthly tele-conferences and plan to hold roughly bi-yearly face-to-face meetings. The first one (Meudon, France in April 2016) was focused on IACT DL3, future meetings will be a bit broader in scope: \ogrameudon
\end{itemize}

Our main work product will be a set of data format specifications for gamma-ray data. Each format usually specifies the names and semantics of data and metadata (a.k.a. ``header'') fields. The scope, status, ongoing discussions, and plans for the data format specifications are presented in the next section. The development of open-source tools and libraries as well as export of existing gamma-ray data to these proposed formats is highly encouraged. However, that work is mainly done by members of the collaborations and software projects mentioned in Figure~\ref{fig:purpose}, who then make suggestions for additions or improvements to the existing specifications.

Currently the process of specification writing is informal and the data format specifications currently written should be seen as proposals, not final standards. We are following the ``release early and often'' philosophy, hoping for feedback and contributions from the larger gamma-ray astronomy community. This approach was motivated by the lack of progress in the past five years on IACT DL3 formats.  Although work has begun within CTA on the development of a DL3 format, CTA doesn't produce DL3 data yet.  Current IACTs were starting to export their data to FITS format and analyzing them with the current science tools, and many slightly different ways to store the same information in FITS files appeared.  Our hope is that this more open format development, making adoption and contributions easy (sending a comment to the mailing list, or making an issue or pull request on Github), will help accelerate the process.  Achieving format stability and dealing with ``requests for enhancement'' after a first stable version of the format specifications is released will be discussed at future meetings.

\begin{figure}[tb]
\centerline{\includegraphics[width=\textwidth]{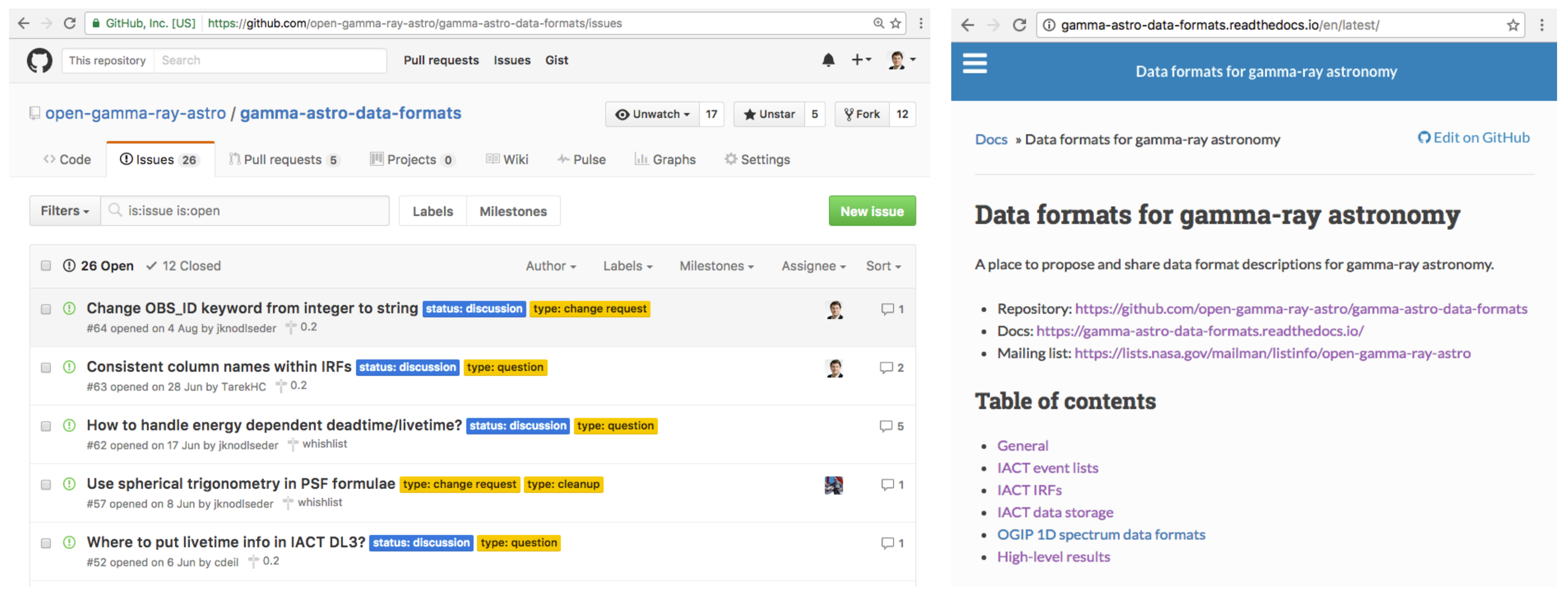}}
\caption{
\emph{Left:} \texttt{gamma-astro-data-formats} Github issue tracker with ongoing discussions. \emph{Right:} latest version of the \texttt{gamma-astro-data-formats} specifications on Read the Docs (PDF and older tagged versions also available).
}
\label{fig:webpage}
\end{figure}

\section{Data models and formats}

This section gives an overview of the current status and plans for the gamma-ray data model and formats. As mentioned before, this effort was only started recently and none of the formats should be considered stable. The next two sections will describe the effort to define an event data model and format (DL3) and higher-level formats for sky-maps, spectra, and lightcurves (DL4), i.e. a content split as already illustrated in Figure~\ref{fig:purpose}.
In the data specification document we have created a ``general'' section where common quantities are defined, such as precise definitions of time scales as well as coordinate systems. 
There are some general topics still under discussion, e.g. there is no consensus on how specific or flexible the format specifications should be. E.g. some people prefer to be very specific (data must be stored in FITS files, data types and units fixed), others would prefer to be flexible (only define header keywords and column names, but data can be stored in other file formats as well, e.g. text-based formats like ECSV).

\subsection{Data level 3 specifications}

\begin{figure}[tb]
\centerline{\includegraphics[width=0.75\textwidth]{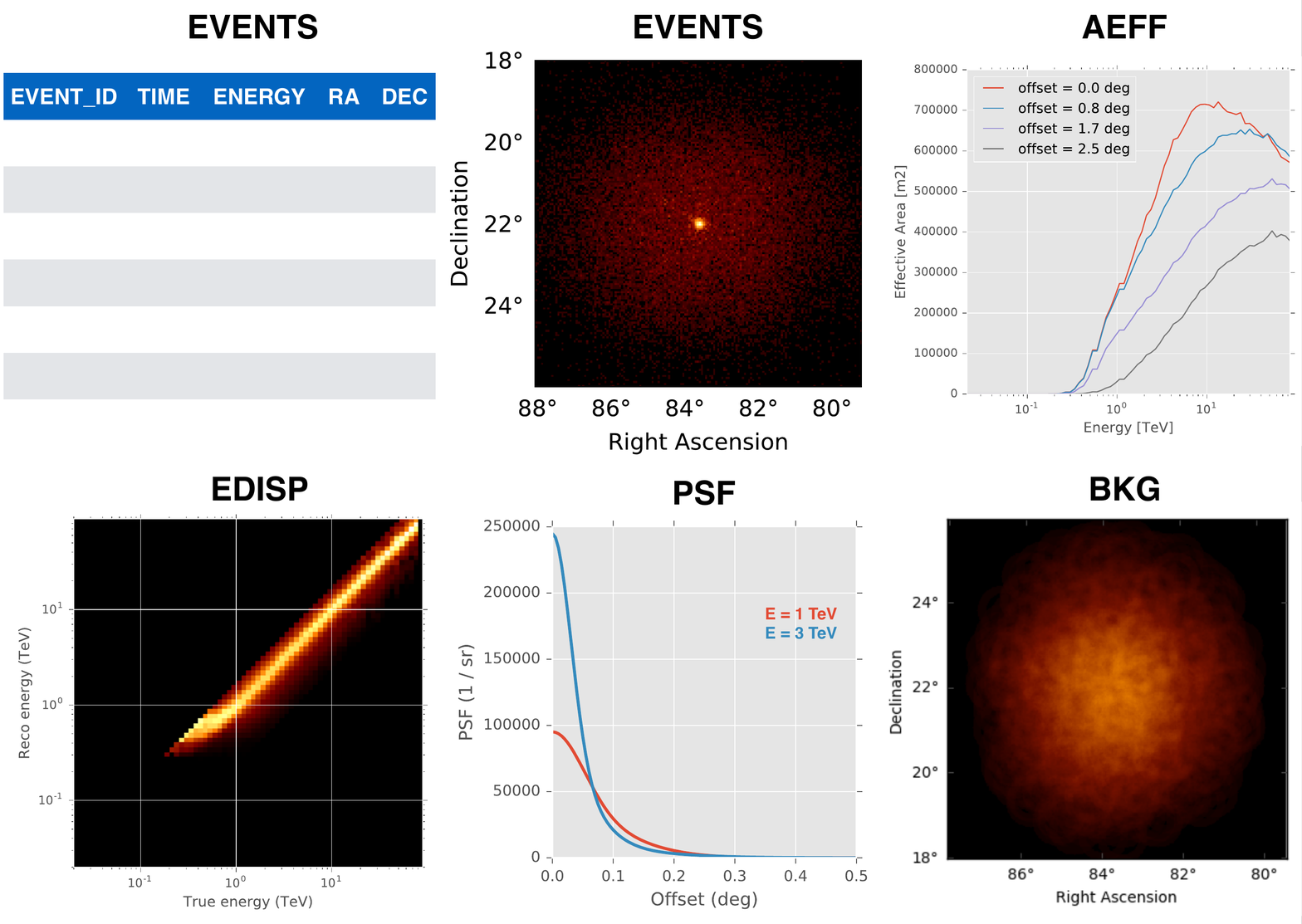}}
\caption{
Illustration of major components of IACT DL3 data (using a H.E.S.S. 1 Crab nebula observation). The \texttt{EVENTS} are stored as a table with the most relevant parameters shown. To derive spectra and morphology measurements of astrophysical sources, instrument response functions (IRFs) are used: effective area (\texttt{AEFF}), energy dispersion (\texttt{EDISP}), and point spread function (\texttt{PSF}). Sometimes background (\texttt{BKG}) models are also created and released as part of DL3 data (as an additional IRF component), and other times they are derived at the science tools level. Note that this picture is not complete, see the ``IACT DL3'' section.
}
\label{fig:iact-dl3}
\end{figure}

The interface between low-level (calibration, shower reconstruction, gamma-hadron separation) and high-level (science tools) analysis for gamma-ray data is usually represented by an event list, where at a minimum the \texttt{EVENT\_ID}, \texttt{TIME}, as well as the reconstructed \texttt{ENERGY} and sky position (\texttt{RA}, \texttt{DEC}) is given for every event. In addition, instrument response functions (IRFs) as well as auxiliary technical information such as telescope configuration options, good time intervals (GTIs), live-time, and pointing information (collectively called \texttt{TECH} in the CTA context) are needed by the science tools to compute exposures, effective resolutions (PSF and EDISP), and ultimately fluxes to compare the data with sky models. This DL3 data, illustrated in Figure~\ref{fig:iact-dl3}, is similar for all gamma-ray telescopes (and other event-recording instruments like e.g. neutrino telescopes). One major difference that affects data formats and analysis tools is whether the gamma-ray telescope was operated in a pointed observation mode (like IACTs most of the time) or in a slewing mode (like HAWC or Fermi-LAT most of the time).

The current specification contains a very preliminary proposal of a data model and formats for IACT DL3 data that is based on ``observations'' (with an \texttt{OBS\_ID}) and assumed stable IRFs during the observation. This proposal was inspired by existing formats used by H.E.S.S. and partly also VERITAS and MAGIC, that are mostly supported by the existing science tool prototypes (Gammapy and ctools). 
A dedicated two-day face-to-face meeting on IACT DL3 data was held in April 2016 in Meudon, France, with 16 participants from all major existing IACTs and CTA.\footnote{\ogrameudon} The use cases and status of efforts to export and archive their data in FITS was presented, as well as the ongoing prototyping in science tools. Many important points were discussed:
\begin{itemize}
\item{}What is an observation? Good time interval? Response time interval?
\item{}How to link \texttt{EVENT} and \texttt{IRF}? (naming conventions, header references, index tables)
\item{}Pointing and live time information
\item{}Exact definition of field of view (FoV) coordinates
\item{}\texttt{IRF} axis specification, validity ranges, errors
\item{}How to support multiple \texttt{EVENT} classes and types?
\end{itemize}
A major result of the face-to-face workshop was to agree to focus on IRF formats that use the multi-array convention and FITS BINTABLE to store the IRF data and axis information, where previously a second format was being developed and prototyped for CTA \citep{2015arXiv150807437W}. The prototyping of IACT DL3 is continuing in the different IACT collaborations and in Gammapy/ctools, with communications online via Github, monthly joint tele-conferences, and a planned face-to-face follow-up meeting in fall 2016. So far the focus is set on pointed gamma-ray observations. Contributions and involvement from people working on slewing telescopes (e.g. Fermi-LAT or HAWC and also IACTs) or non-gamma-ray telescopes with similar data (e.g. neutrino telescopes) are welcome. The largest stakeholder for the IACT DL3 work is CTA.

\subsection{Data level 4 \& 5 specifications}

\begin{figure}[tb]
\centerline{\includegraphics[width=0.8\textwidth]{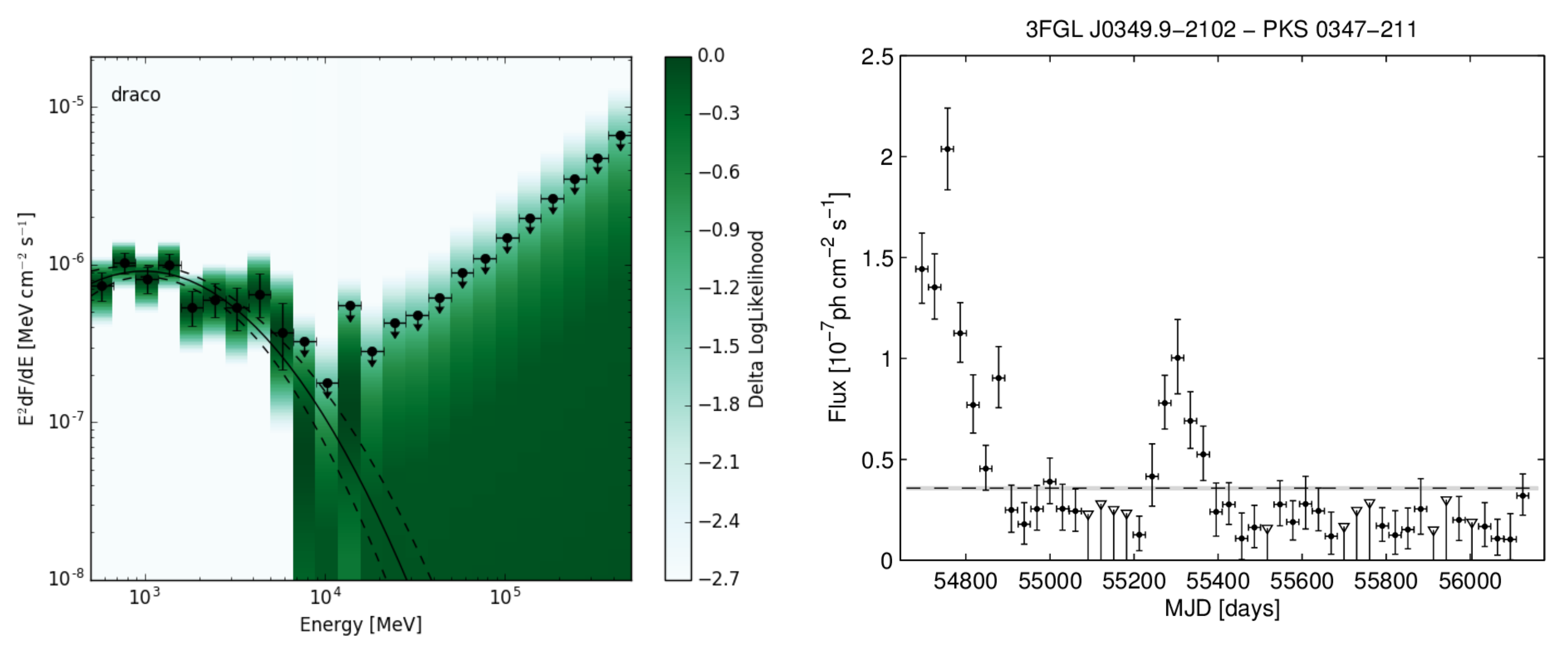}}
\caption{
Gamma-ray ``data level 4'' examples. \emph{Left:} spectral energy distribution (SED) likelihood profiles (green), with flux points, upper limits and best-fit model shown. \emph{Right:} Lightcurve of 3FGL~J0349.9-2102 from the third Fermi-LAT catalog.
}
\label{fig:dl4-examples}
\end{figure}

Another topic in the \texttt{gamma-astro-data-formats} specifications is the development of formats to store high-level data products such as sky-maps, spectra, and lightcurves (data level 4) or source catalogs (data level 5).  Here we list DL4 and DL5 format specifications that are currently included or under consideration:
\begin{itemize}
\item{} For 2-dimensional images, the existing FITS and world coordinate system (WCS) standard provides a solution that works for gamma-ray sky-maps as well. If something gamma-ray specific were to be added, it would likely be specifications on how to store parameters of interest for analysis or provenance in the header.
\item{} For 3-dimensional cubes, where the third dimension is \texttt{ENERGY}, commonly 3-dimensional \texttt{FITS IMAGE} extensions are used. However, due to either the complexity or missing features in the FITS WCS model, the energy axis information is not represented in the FITS header, but in a separate \texttt{BINTABLE HDU} called \texttt{ENERGY} (if the cube represents quantities at given energies, like exposure or flux), or \texttt{EBOUNDS} ("energy bounds", if the cube represents integral quantities like e.g. counts).
A specification at \gadf can document the exact semantics for storing the energy axis and how interpolation and integration should be performed by science tools (e.g. for exposure or diffuse model flux cubes).
\item{} For all-sky maps and cubes, HEALPix\cite{2005ApJ...622..759G} is commonly used in gamma-ray astronomy (e.g. by Fermi-LAT). While 2-dimensional HEALPix images are standardized, extensions have been developed to represent cubes, as well as to store sparse data or images that don't cover the whole sky \footnote{\pointlikedata}. These gamma-ray specific extensions are not standardized, and a specification at \gadf would be welcome.
\item{} The common method for 1-dimensional spectral analysis in X-ray astronomy \citep{Davis:2001}, as well as the corresponding file formats (e.g. \texttt{ARF} for effective area, \texttt{RMF} for energy dispersion) are also used in VHE gamma-ray astronomy. In the current specification we have added a section referencing the relevant OGIP documents and explained how the formats are commonly used in gamma-ray astronomy (e.g. using a ``reconstructed energy'' axis instead of the ``pulse height channels'' axis used in X-ray astronomy).
\item{} For 1-dimensional spectra, a format to store flux points and upper limits, as well as full likelihood profiles, is available at \gadf (see Figure~\ref{fig:dl4-examples} left panel). It was first developed in Fermipy and applied to Fermi-LAT analyses, and is now being adopted for IACT spectra.
\item{} No format specification for light curves (see Figure~\ref{fig:dl4-examples} right panel for an illustration) is available yet. Previously a format has been proposed in \cite{2010AnA...524A..48T} and a pull request with discussions for a lightcurve specification at \gadf is ongoing.
\item{} No format specifications have been proposed for catalogs (data level 5, DL5) yet. So far each catalog (Fermi-LAT, upcoming H.E.S.S. and HAWC) is unique (but all similar) and some science tools have per-catalog code to produce corresponding sky models.
\end{itemize}

\section{Conclusions}

In early 2016, we have started the \gadf effort to create an open forum (mailing list, Github, meetings) and eventually open and common data formats for space- and ground-based gamma-ray instruments. This effort is similar to the HEASARC FITS working group from the 1990s, but this time driven mainly by the movement of ground-based gamma-ray observatories toward producing high-level gamma-ray data in FITS format (IACT DL3 data). We invite everyone interested in this topic to join the mailing list, regular meetings and to contribute or give feedback on how the current formats could be improved to support your use cases.

\section{Acknowledgements}

We would like to thank everyone that has contributed to or supported this
effort, be it directly via contributions to the format specification, or
indirectly via feedback or adopting the existing formats, spending the
effort to transform their existing data to the common formats defined here, or by giving people time or travel money to work on this.

We would also like to thank the following services: \texttt{NASA} for hosting the \texttt{open-gamma-ray-astro} mailing list, \texttt{Github} for making this way of online collaboration possible, \texttt{Sphinx} as documentation system and \texttt{Read the docs} for building and hosting the HTML and PDF version of the specification.


\nocite{*}
\bibliographystyle{aipnum-cp}
\bibliography{open-gamma-ray-astro-gamma2016}

\end{document}